\documentstyle[11pt,graphicx]{article}

\newcommand{\be}{\begin{equation}}
\newcommand{\ee}{\end{equation}}
\newcommand{\ba}{\begin{eqnarray}}
\newcommand{\ea}{\end{eqnarray}}

\begin{document}

\title{Superconductivity as a Bose-Einstein condensation?}

\author{
S.K. Adhikari$^{\mbox{\scriptsize a}}$, M. Casas$^{\mbox{\scriptsize b}}$,  A. Puente,$^{\mbox{\scriptsize b}}$ A. Rigo,$^{\mbox{\scriptsize b}}$
M. Fortes,$^{\mbox{\scriptsize c}}$ \\ 
M.A. Sol\'{\i}s,$^{\mbox{\scriptsize c}}$
M. de Llano,$^{\mbox{\scriptsize d}}$ A.A. Valladares,$^{\mbox{\scriptsize d}}$ and O. Rojo$^{\mbox{\scriptsize e}}$ 
}

\date{}

\maketitle

\noindent $^{\mbox{\scriptsize a}}${Instituto de F\'{\i}sica Te\'{o}rica, Universidade Estadual 
Paulista, 01405-900 S\~{a}o Paulo, SP, Brazil}\\
$^{\mbox{\scriptsize b}}$Deparment de F\'{\i}sica, Universitat de les Illes Balears, 07071 Palma de Mallorca, Spain \\
$^{\mbox{\scriptsize c}}${Instituto de F\'{\i}sica, Universidad Nacional Aut\'{o}noma de 
M\'{e}xico, \\ Apdo. Postal 20-364, 01000 M\'{e}xico, DF, M\'{e}xico} \\
$^{\mbox{\scriptsize d}}${Instituto de Investigaciones en Materiales, Universidad Nacional
Aut\'{o}noma de M\'{e}xico, \\ Apdo. Postal 70-360, 04510 M\'{e}xico, DF,
M\'{e}xico}\\
$^{\mbox{\scriptsize e}}${PESTIC, Secretar\'{\i}a Acad\'{e}mica \& CINVESTAV, IPN, 04430 M\'{e}xico DF,
M\'{e}xico}

\begin{abstract}
Bose-Einstein condensation (BEC) in two dimensions (2D) (e.g., to describe the
quasi-2D cuprates) is suggested as the possible mechanism
widely believed to underlie superconductivity in general.  A
crucial role is played by nonzero center-of-mass momentum Cooper pairs
(CPs) usually
neglected in BCS theory.  Also vital is the unique {\it linear} dispersion
relation appropriate to weakly-coupled ``bosonic" CPs moving in the Fermi
sea---rather than in vacuum where the dispersion would be quadratic but only
for very strong coupling, and for which BEC is known to be impossible in 2D. \\

{\bf Corresponding author: }\ \ M. de Llano dellano@servidor.unam.mx \\
\end{abstract}

\maketitle
Bose-Einstein condensation (BEC) of Cooper pairs (CPs) leads to a phase
transition even in 2D in any many-fermion system dynamically capable of
forming CPs.
This transition could be germane to superconductivity in the quasi-2D
cuprates.
In the weak coupling limit one finds a
nearly linear dispersion relation for the CP that suggests very high, even
diverging, critical temperatures $T_{c}$. On the other hand, in the strong
coupling
limit a nearly quadratic dispersion relation gives vanishingly small $T_{c}$'s.
For intermediate coupling one gets the finite $T_{c}$'s
appropriate for
real quasi-2D superconductors.

The single-Cooper pair problem may appear academic at first but has
significant consequences. The familiar BEC formula for the transition temperature
$T_{c}\simeq 3.31\hbar ^{2}n_{B}^{2/3}/m_{B}k_{B}$, with $n_{B}$ the number
density of bosons of mass $m_{B}$ and $k_{B}$ the  Boltzmann constant, is a
special case of the more general expression \cite{pla} valid for any space
dimensionality $d>0$ and any boson dispersion relation $\varepsilon_{K}=C_{s}\,K^{s}$ with $s > 0$ and $C_{s}$ 
a constant, given by
\begin{equation}
T_{c}=\frac{C_{s}}{k_{B}}\left[ \frac{s\,\Gamma (d/2)\,(2\pi )^{d}n_{B}}{2\pi
^{d/2}\,\Gamma (d/s)g_{d/s}(1)}\right] ^{s/d}.  
\label{gentc}
\end{equation}
If $\mu _{B}(T)$ is the boson
chemical potential and $e^{\mu _{B}(T)/k_{B}T} \equiv z$ the fugacity,
$g_{\sigma}(z)\equiv \sum_{l=1}^{\infty }z^{l}/l^{\sigma}$.
For $z=1$ and $\sigma \geq 1$ this is just $\zeta (\sigma)$, the Riemann
Zeta-function of order $\sigma $ which is finite for $\sigma > 1$ and infinite
for $\sigma = 1$, while the series $g_{\sigma }(1)$ diverges for
all $\sigma \leq 1$. \ For $s=2$, $C_{2}=\hbar ^{2}/2m_{B}$, and since $\zeta
(3/2)\simeq 2.612$, this leads to the BEC $T_{c}$-formula cited above.
\ Since $g_{d/2 }(1)$ diverges for all $d/2 \leq 1$, $T_{c}=0$
for all $d\leq 2$.  Eq. (\ref {gentc}) follows from the boson number equation
\begin{equation}
N=N_{0}(T)+\sum_{K\neq 0}\left[ e^{\{\varepsilon _{K}-\mu
_{B}(T)\}/k_{B}T}-1\right] ^{-1}  
\label{bec}
\end{equation}
where $N_{0}(T)$ is the number of bosons in the $K=0$ state.
At $T=T_c$ both $N_0(T_c)$ {\it and} $\mu_B(T_c)$ virtually vanish so
that using $\sum_{k} \rightarrow \left( L/2\pi \right) ^{d}\int d^{d}k$
in (\ref{bec}) then yields (\ref {gentc}).

The fact that a CP can have a {\it linear}, as opposed to the usual
quadratic,
dispersion relation was mentioned as far back as 1964 by Schrieffer
\cite{sch64}, p. 33.
We have found it to be so for weak coupling; it becomes
quadratic only for extremely strong coupling.  In weak coupling
$\varepsilon _{K}\equiv \Delta _{0}-\Delta _{K} \simeq a(d)\hbar v_{F}K$, where
$\Delta _{K}$ is the (positive) binding energy of a CP of center-of-mass
momenta (CMM)
$\hbar K$, $v_{F}\equiv \hbar k_{F}/m$ is
the Fermi velocity, $m$ the fermion effective mass,
while \  $a(d)\equiv $ $2/\pi$ and $1/2$
in 2D and 3D, respectively \cite{physc}. For linear dispersion $s=1$,
$C_{1}=$ $a(d)\hbar v_{F}$, another special case of
(\ref{gentc}) is
\begin{equation}
T_{c}={\frac{a(d)\hbar v_{F}}{k_{B}}}\left[ {\frac{\pi ^{({d+1)/2}}\,n_{B}}{
\Gamma \lbrack ({d+1)/2]g}_{d}(1{)}}}\right] ^{1/d}.  \label{tcs1}
\end{equation}
Now $T_{c}=0$ for all $d \leq 1$ only---and $T_{c}>0$ for all $\ d>1$, which
is precisely the range of dimensionalities for all known superconductors if one
includes the quasi-1D organo-metallic Bechgaard salts \cite{quai1d}. Using
the interpolation $a(d) = (7/2-6/\pi)+(8/\pi-13/4)d+(3/4-2/\pi)d^2$, which
reduces to 1, $2/\pi$ and $1/2$ in 1D, 2D and 3D, respectively, Fig.
1 graphs
(\ref{gentc}) and (\ref{tcs1}) {\it vs} $d$ if one imagines all fermions in the
many-fermion system paired into CPs of mass $m_{B}=2m$.  The particle number
density of the original fermions is $n \equiv k^{d}_{F}/2^{d - 2} \pi^{d/2}
d\;\Gamma (d / 2)= 2n_{B}$.  Here
$k_{B}T_{F}
\equiv E_{F}= \hbar^2 k_F^2/2m$. These curves are an {\it upper bound}
to a more
realistic $T_c$ where only a
fraction of all fermions are actually bound.

\begin{figure}[htb]
\begin{center}
\includegraphics[width=10cm,height=10cm,angle=0]{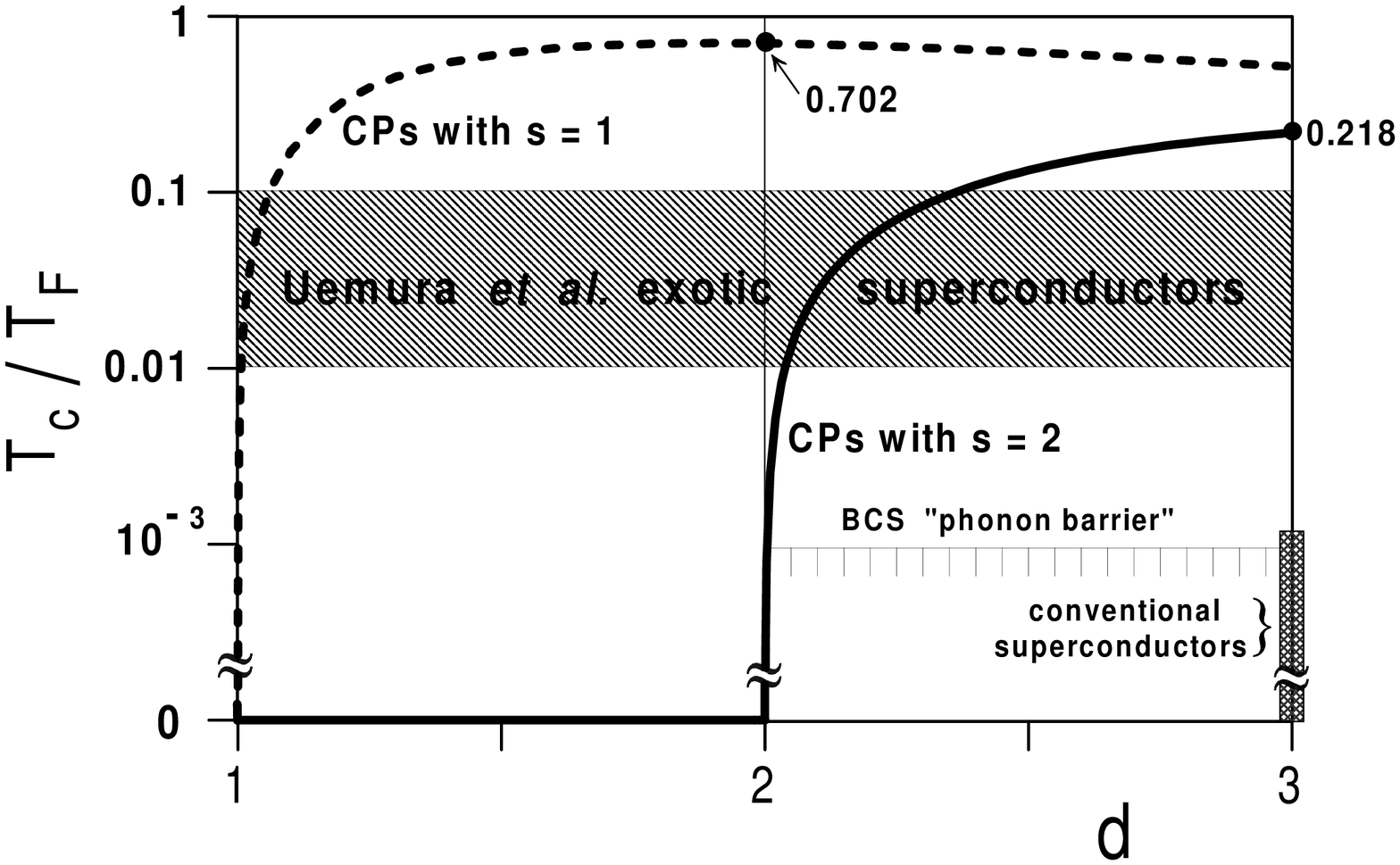}
\vspace{-2.cm}
\begin{quotation}
\caption{Upper bounds to BEC critical temperatures $T_{c}$ as function of
system
dimensionality $d$, as explained in text.  Shaded areas refer to empirical data \cite{Ue} while BCS
\lq\lq phonon barrier" refers to BCS $T _{c} \simeq 1.13 \Theta_{D} \exp({-1/\lambda})$ with
$\Theta _{D}$ the Debye
temperature $\simeq 300$ K and  $\lambda \equiv N(0)V =1/2$ where $
N(0)$ is the density of states at the Fermi surface and $V$ the attractive BCS 
coupling constant.}
\end{quotation}
\end{center}
\end{figure}

The above holds for ``unbreakable" CPs, i.e., bosons existing for all CMM so
that the integrals over momenta go from 0 to $\infty$.
However, CPs actually break up \cite{physc} into two ``unpaired fermions" above a
certain CMM $K_{0}$.  This increases
in value as coupling is increased and tends to infinity (as will be seen
below) for
infinite coupling.  The CP breakup $K_{0}$, defined by $\Delta _{K_{0}}=0$,
for
very weak coupling is just $K_{0}\simeq \Delta_{0}/a(d)\hbar
v_{F}\rightarrow 0$
as $\Delta_{0}\rightarrow 0$.  Eq. (\ref {bec}) with $K_{0}$ rather than
$\infty$
as an upper limit eventually yields
\begin{equation}
T_c \simeq \frac{(d-1)[2\pi a(d) \hbar v_F]^d n_{B}}{A_d k_B \Delta_0^{d-1}}
\mathrel{\mathop{\longrightarrow}\limits_{\Delta_0
\rightarrow 0}}
\infty\, ,
\label{tcinfty}
\end{equation}
where $A_d=2\pi$ or $4\pi/3$ in 2D or 3D, respectively. Hence, since in
that limit any remaining CPs are condensed in the $K=0$ state at all
temperatures, the critical temperature $T_{c}$ must be {\it infinite}.
Note that $T_{c}\equiv 0$ explicitly for $d=1$.  Indeed, if
the unpaired fermions resulting from broken CPs are taken into account in a
more
realistic {\it binary boson-fermion mixture model} this artificially infinite $T_{c}$
is ``tamed" down \cite{pla}\ to a finite $T_{c}$---which remarkably enough
practically coincides with that of the pure unbreakable boson gas
(\ref{tcs1}).
Detailed calculations using the BCS model interaction gives
a $T_c$ still almost an order of magnitude too large compared with
the empirical range \cite{Ue} of $T_{c}\approx (0.01-0.1)T_{F}$ for
cuprates---even for a moderate coupling of $\lambda =1/2$.  However,
the calculated $
T_{c}$ of about 800 K is about 17 times larger than that of the BCS theory
$T_{c} \simeq 1.13\Theta_{D} \exp(-1/\lambda)$ of about 46 K.  For
cuprates $d \simeq 2.03$ has been suggested \cite{wk}\ as more realistic as
it accounts for inter-layer couplings, but results would be very similar to
those discussed here for $d=2$.

A far more general interfermion interaction is the $S$-wave
attractive
separable potential \cite{18a} whose double Fourier transform is
\begin{equation}
V_{pq}=-(v_{0}/L^{2})g_{p}g_{q}.  \label{1}
\end{equation}
Here $L$ is the size of the ``box" confining the many-fermion system,
$v_{0}\geq 0$ is the interaction strength and $g_{_{p}} \equiv
(1+p^{2}/p_{0}^{2})^{-1/2}$
where $p_{0}$ is the inverse range of the potential.  Hence,
e.g., $p_{0}\rightarrow \infty $ implies $g_{_{p}}=1$ and corresponds to
the contact or delta potential $V(r)=-v_{0}\delta ({\bf r})$, and
$p_{0}=k_{F}$ to
a range of order of the average interfermion spacing, etc. Such an
interaction
model may mimic a wide variety of possible dynamical mechanisms in
superconductors:
a force mediated by phonons, or plasmons, or excitons, or magnons, etc., or
even
a purely electronic interaction.  In the first instance mentioned one may
have, e.g., a (possibly singular) coulombic interfermion repulsion
surrounded by
a longer-ranged electron-phonon attraction. The former
is central to the problem at hand; the latter (whether or not phonon-based)
is indispensable to create the CPs detected experimentally
\cite{fq} in both conventional as well as cuprate superconductors.  In vacuum,
a two-body bound state with (positive) binding energy $B_{2}\geq 0$ is
possible for any 2D potential with sufficient attraction.  The vacuum
$t$-matrix
\cite{AD} associated with the Lippmann-Schwinger equation then develops a
pole at an energy $E=-B_{2}$ so that for any dimension $d$ (\ref{1}) leads to
\begin{equation} \frac{L^{d}}{v_{0}}=\sum_{k}\frac{g_{k}^{2}}{B_{2}+\hbar
^{2}k^{2}/m} .\;\quad \;\;\; \label{4} \end{equation}
In 1D the integral on the rhs is exact and gives the well-known binding energy
\ $B_{2}=mv_{0}^{2}/4\hbar ^{2}$ of the single bound state, but in 2D that
integral diverges logarithmically for large $k$.
On the other hand, the CP equation for two fermions above the Fermi surface
(of energy $E_F$)
with momenta wavevectors ${\bf k}_{1}$ and ${\bf k}_{2}$ (and CMM
wavevector ${\bf K}\equiv {\bf k}_{1}+{\bf k}_{2}$) is given by
\begin{equation}
\sum_{k}{}^{^{\prime }}{}\frac{g_{k}^{2}}{\hbar ^{2}k^{2}/m+\Delta
_{K}-2E_{F}+\hbar ^{2}K^{2}/4m}=\frac{L^{d}}{v_{0}}  \label{7}
\end{equation}
where ${\bf k}\equiv \frac{1}{2}({\bf k}_{1}-{\bf
k}_{2})$
is the relative momentum wavevector, $2E_{F}-\Delta _{K}$ the
total pair energy, $\Delta _{K}\geq 0$ as before, and the prime on
the summation implies restriction to states {\it above} the Fermi
surface, viz., $|{\bf k}\pm {\bf K}/2|>k_{F}$.
In principle, (\ref{7}) can be solved
numerically for $\Delta _{K}$ as a function of coupling strength $v_{0}$
for all
$0\leq v_{0}<+\infty$.  More conveniently, however, for an interaction
of the form (\ref{1}) one can solve for $\Delta _{0}$ (analytically
if  $g_k=1$) and for $\Delta _{K}$ numerically, both as functions
of $B_{2}$ alone, after eliminating the interaction strength $v_{0}$ by
combining (\ref{4}) and (\ref {7}).  This yields a {\it renormalized CP
equation} employing $0\leq B_{2}/E_{F}<\infty $ \cite{miyake} as dimensionless
interaction coupling parameter instead
of the (possibly singular) potential strength term $1/v_{0}$ is now
eliminated.  A much greater generality for the
equation with $g_{k}=1$, viz., for
{\it any} interfermion interaction with binding energy $B_{2}$, is conceivable. 
In 2D one has the
surprising result (traceable to the fact that unlike 1D or 3D the
fermionic density of states is a {\it constant} independent of $k_{F}$)
that $\Delta_{0}=B_{2}$ for {\it all} coupling, at
least for an
attractive delta interaction.  Fig. 2 shows exact numerical results for the nonnegative {\it CP excitation
energy} $\varepsilon _{K}\equiv \Delta _{0}-\Delta _{K}$ mentioned earlier,
 for zero range (full curves).  For very
weak coupling the exact curves are virtually linear, i.e.,
$\varepsilon _{K} \rightarrow 2\hbar v _{F} K/\pi$,
and for very
large coupling tend aymptotically to
the exact quadratic $\hbar^2 K^2/ 2(2m)$ (short-dashed curves).  The
long-dashed loops (reminiscent of the ``roton" part of the excitation spectrum of
liquid $^4$He) emerge for finite range at sufficiently strong coupling.

\begin{figure}[htb]
\begin{center}
\includegraphics[width=10cm,height=11cm,angle=0]{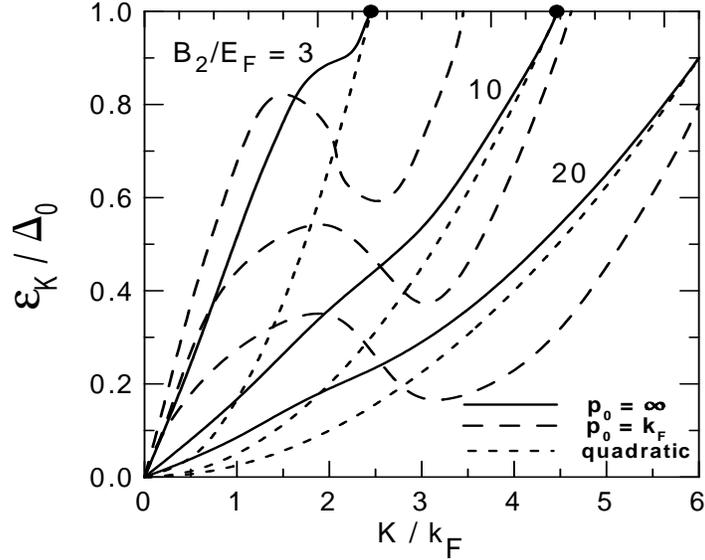}
\vspace{-3.cm}
\begin{quotation}
\caption{Dimensionless CP excitation energy $\varepsilon _{K}/\Delta _{0}$
{\it vs}
dimensionless CMM wavenumber $K/k _{F}$, as explained
in text. Full curves are exact zero-range results; short-dashed the quadratic
approximation; and long-dashed the exact finite-range result with
$p_{0}=k_{F}$. Dots mark CP breakup points.
}
\end{quotation}
\end{center}
\end{figure}

When the zero-range results like those in Fig. 2 are used in (\ref {bec})
assuming $n_{B}=n/2$ and $m_{B}=2m$, one
gets Fig. 3 for the BEC $T_{c}$'s of a pure gas of breakable or unbreakable
CPs. Consequently, for a substantial range of intermediate couplings in 2D {\it
finite}
BEC $T_{c}$'s are possible that lie within the empirical range \cite{Ue}
of cuprate
$T_{c}$'s even for a simple pure boson gas of CPs. \ Significantly,
$T_{c}$ is {\it no longer} zero as would be predicted in a BEC picture by a
quadratic relation appropriate for ``local-pair" CPs in vacuum---a result that
convinced many that BEC is irrelevant for quasi-2D cuprate superconductors.
More
accurate BEC $T_{c}$'s should include refinements such as allowing for {\it
unpaired
fermions } in a more realistic binary boson-fermion mixture model, CP-fermion
interactions, non-$S$-wave interactions, etc.

\begin{figure}[htb]
\begin{center}
\includegraphics[width=10cm,height=11cm,angle=0]{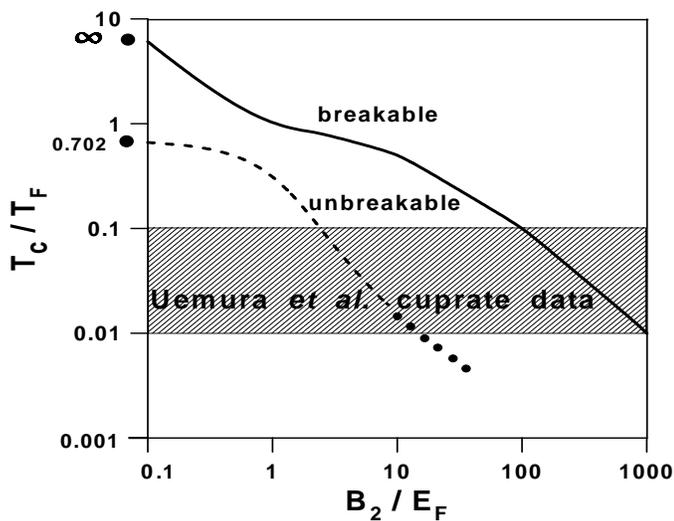}
\vspace{-3.cm}
\begin{quotation}
\caption{Critical BEC $T_{c}$ (in units of $T_{F}$) calculated from (\ref{bec})
for a pure
gas of breakable (i.e., with finite limit $K_{0}$) 
CPs using the exact numerical dispersion illustrated in Fig. 2,
{\it vs} all coupling  $0\leq B_{2}/E_{F}<\infty$. The unbreakable (i.e., $K_0 = \infty$) case corresponds to the interpolation {\it ansatz} $\varepsilon_K = \frac{2}{\pi}\hbar v_F K [1-\mbox{tanh}(B_2/E_F)]+ \frac{\hbar^2 K^2}{2(2m)}\mbox{tanh}(B_2/E_F)$.}
\end{quotation}
\end{center}
\end{figure}
Finally, CPs are here considered ``bosonic'' even though they do {\it not} obey
(Ref. \cite{sch64} p. 38) Bose commutation relations. This is because for a
given $K$ they have {\it indefinite} occupation number as (in the
thermodynamic
limit) there are an indefinitely large number of allowed
(relative wavevector) ${\bf k}$ values corresponding to an indefinitely large number
of possible pairs of vectors ${\bf k}_{1}$ and ${\bf k}_{2}$.  Hence, for
any coupling
and thus {\it any degree of overlap} between them, CPs do in fact obey the
BE distribution (\ref{bec}) from which BEC is determined.

To conclude, the single CP problem with non-zero CMM (usually neglected in
BCS theory) is illustrated in 2D for all coupling (and/or normal-state fermion
density).  BEC then suggests itself as
a possible mechanism for superconductivity---in fact, for all
observed superconductors from the quasi-1D organo-metallics to the more
familiar 3D ones.  But one must employ the correct dispersion relation in the
CMM of the CPs.
This relation is complicated but obtainable numerically for arbitrary
coupling, and is purely linear for very weak and purely quadratic
for very strong coupling. \\

Partial support from UNAM-DGAPA-PAPIIT
(M\'{e}xico) \#
IN102198, CONACyT (M\'{e}xico) \# 27828 E, DGES (Spain) \# PB95-0492 and
FAPESP\ (Brazil) is acknowledged.

\end{document}